\documentclass[prd,showpacs,twocolumn,superscriptaddress,nofootinbib,floatfix]{revtex4-2}
\usepackage{setspace}
\usepackage[utf8]{inputenc}
\usepackage{makecell}

\usepackage{orcidlink}
\usepackage{float}
\usepackage{amsmath,amssymb,amsfonts,bm}
\usepackage{graphicx}
\usepackage{multirow}
\usepackage{xcolor}
\definecolor{poscolor} {RGB} {252,188,190} 
\definecolor{negcolor} {RGB} {168,168,234} 
\usepackage{slashed}
\usepackage[normalem]{ulem}
\usepackage{amsmath}
\usepackage{txfonts}
\usepackage{amssymb}
\usepackage{slashed}
\usepackage{color}
\usepackage{multirow}
\usepackage{float}
\usepackage{graphicx,color,dcolumn,bm}
\usepackage{subfigure}

\usepackage{hyperref}
\hypersetup{colorlinks,citecolor=purple,anchorcolor=red,menucolor=red, linkcolor=red,filecolor=red,runcolor=red,urlcolor=blue,frenchlinks=true}
\usepackage{appendix}

\allowdisplaybreaks

\newcommand{\pfield}{\mathcal{P}}
\newcommand{\pbfield}{\mathcal{\bar P}}

\newcommand{\xju}{\affiliation{School of Physics Science and Technology, Xinjiang University, Urumqi, Xinjiang 830046 China}}
\newcommand{\bonn}{\affiliation{Helmholtz Institut f\"{u}r Strahlen- und Kernphysik and Bethe Center for Theoretical Physics,\\ 
Universit\"{a}t Bonn, D-53115 Bonn, Germany}}
\newcommand{\fzj}{\affiliation{Institute for Advanced Simulation (IAS-4), Forschungszentrum J\"ulich, D-52425 J\"ulich, Germany}}
\newcommand{\peng}{\affiliation{Peng Huanwu Collaborative 
Center for Research and Education, Beihang University, Beijing 
100191, China}}
\begin{document}
\title{Role of the short-range dynamics in the formation of $D^{(*)}\bar D^{(*)}/B^{(*)}\bar B^{(*)} $ hadronic molecules}

\author{Nijiati Yalikun\orcidlink{0000-0002-3585-1863}}
\email{nijiati@xju.edu.cn}
\xju
\author{Xiang-Kun Dong\orcidlink{0000-0001-6392-7143}}
 \email{xiangkun@hiskp.uni-bonn.de}
\bonn

\author{Ulf-G. Mei{\ss}ner\orcidlink{0000-0003-1254-442X}}\email{meissner@hiskp.uni-bonn.de}
\bonn\fzj\peng


\begin{abstract}
 We investigate potential hadronic molecular states in the $D^{(*)}\bar D^{(*)}$ and $B^{(*)}\bar B^{*}$ systems using light meson exchange interactions. Our analysis focuses on coupled-channel systems with spin-parity quantum numbers $J^{PC}=0^{++}$, $1^{+\pm}$, and $2^{++}$, examining how the $\delta(\bm r)$ potential affects states near threshold. Using coupled-channel analysis, we reproduce the $X(3872)$ mass with a given cutoff for the $(I)J^{PC}=(0)1^{++}$ state, finding a minimal impact from the $\delta(\bm r)$ term. At this cutoff, both the $(0)0^{++}$ state near the $D\bar D$ threshold and the $(0)1^{+-}$ state near the $D\bar D^*$ threshold show less sensitivity to the $\delta(\bm r)$ term compared to the three states—$(0)0^{++}$, $(0)1^{+-}$, and $(0)2^{++}$—near the $D^*\bar D^*$ threshold. As anticipated, the $B^{(*)}\bar B^{*}$ systems exhibit similar behavior but with stronger binding due to their larger reduced mass. These findings suggest promising directions for future experimental searches, particularly in the isoscalar sector, which could substantially advance our understanding of exotic tetraquark states.
\end{abstract}

\maketitle

\newpage 

\section{Introduction}~\label{sec:1}
The formation mechanism of the hadrons governed by the strong interaction described by quantum chromodynamics (QCD) is intensively studied, however, its low-energy dynamics is still the most challenging task for the hadron community. It is no clear that apart from the well-known $qqq$ baryons and $q\bar q$ mesons of the conventional quark model~\cite{GellMann:1964nj,Zweig:1964jf}, there exist multiquark states, glueballs, quark-gluon hybrids, which are collectively called exotic hadrons. Multiquark states can be classified into tetraquark states ($qq\bar q \bar q$), pentaquark states ($qqq q\bar q$), and so on. The study of multiquark states, especially how the quarks are grouped inside (i.e., compact or molecular configuration) plays a crucial role for understanding the structure formation in QCD; for a few recent reviews, see Refs.~\cite{Lebed:2016hpi,Esposito:2016noz,Guo:2017jvc,Liu:2019zoy,Chen:2022asf,Hanhart:2025bun}.

As the most famous exotic state, the $X(3872)$, listed as $\chi_{c1}(3872)$ in the Review of Particle Physics (RPP)~\cite{ParticleDataGroup:2024cfk}, was observed by the Belle collaboration two decades ago~\cite{Belle:2003nnu}, whose mass is very close to the $D^0\bar D^{*0}$ threshold. Due to its quantum numbers $J^{PC}=1^{++}$~\cite{LHCb:2013kgk} and large isospin breaking in decays~\cite{LHCb:2022jez,Dias:2024zfh}, it has been extensively discussed as a potential $D\bar D^*+\rm{c.c.}$ hadronic molecule~\cite{Barnes:2003vb,Voloshin:2003nt,Tornqvist:2004qy, Suzuki:2005ha,AlFiky:2005jd, Hanhart:2007yq,Thomas:2008ja,Fleming:2008yn,Liu:2008fh, Gamermann:2009fv,Bignamini:2009sk, Nieves:2012tt,Guo:2013zbw,Voloshin:2013dpa,Wang:2013vex,Prelovsek:2013cra,Hidalgo-Duque:2013pva,Swanson:2014tra,Albaladejo:2015lob,Baru:2015nea,Chen:2016qju,Albaladejo:2017blx,Guo:2019qcn,Voloshin:2019ivc,Yang:2020nrt,Zhang:2020mpi,Dong:2020hxe,Dong:2021bvy, Du:2022jjv}. While other interpretations—such as compact tetraquark configurations or mixed molecular-charmonium states—have also been proposed~\cite{Wang:2023ovj,Kalashnikova:2005ui,Barnes:2007xu,Ortega:2009hj,Esposito:2021vhu}, a recent analysis of the most precise experimental data concludes that the compositeness of the $X(3872)$ is consistent with unity, strongly supporting its interpretation as a pure molecular state~\cite{Ji:2025hjw}. Following the discovery of the $X(3872)$, many charmonium- and bottomonium-like states are reported experimentally, $Z_c(3900)$~\cite{BESIII:2013ris,Belle:2013yex,BESIII:2013qmu}, $Z_c(4020)$/$Z_c(4025)$ ~\cite{BESIII:2013mhi,BESIII:2013ouc,Belle:2014nuw,BESIII:2014gnk}, $Z_b(10610)$ and $Z_b(10650)$~\cite{Belle:2011aa,Belle:2013urd,Belle:2014vzn}, and assigned with isospin $1$. These states are close to the $D\bar D^*$, $D^*\bar D^*$, $B\bar B^*$, $B^*\bar B^*$ thresholds, respectively. Conspicuously, their isoscalar partners and the states near thresholds of $D\bar D$/ $B\bar B$ are absent from the list. Apart from that, the observation of hidden-charm pentaquark states $P_c$~\cite{LHCb:2015yax,LHCb:2019kea} and $P_{cs}$~\cite{LHCb:2020jpq,LHCb:2022jad} by the LHCb collaboration adds further members to the exotic hadron zoo. They can be understood as hadronic molecules in $\bar D^{(*)} \Sigma_c^{(*)}$ and $\bar D^{(*)}\Xi_c^{(*)}$ system~\cite{Dong:2021juy,Karliner:2022erb,Wang:2022mxy,Yan:2022wuz,Meng:2022wgl,Yang:2022ezl,Liu:2020hcv,Chen:2020kco,Wang:2020eep,Peng:2020hql,Chen:2020uif,Du:2021bgb,Xiao:2021rgp,Feijoo:2022rxf,Nakamura:2022jpd,Yan:2022wuz}. A very intriguing and remarkable fact is that most of the exotic states are located quite close to the thresholds of a pair of hadrons that they can couple to. This property can be understood as there is an $S$-wave attraction between the relevant hadron pair~\cite{Dong:2020hxe}, and it naturally leads to the hadronic molecule interpretation~\cite{Chen:2016qju,Guo:2017jvc,Brambilla:2019esw,Yamaguchi:2019vea,Dong:2021juy,Dong:2021bvy}. The validity of the hadronic molecule picture is also reflected by the successful quantitative predictions of some exotic states in early theoretical works based on the hadron-hadron  interactions~\cite{Tornqvist:1993ng,Wu:2010jy,Wu:2010vk,Wang:2011rga,Yang:2011wz,Wu:2012md,Xiao:2013yca,Uchino:2015uha,Karliner:2015ina}.

In present work, we solve the stationary Schr\"odinger equation to search for possible molecular states in $D^{(*)}\bar D^{(*)}$ and $B^{(*)}\bar B^{*} $ systems, similar to the strategies used previously in the hidden-charm pentaquark sector~\cite{Yalikun:2021dpk,Yalikun:2021bfm}. The low-energy interaction between hadrons is described by the one-boson-exchange model (OBE), which includes the SU(3) vector-nonet mesons, pseudoscalar octet mesons, and the $\sigma$ meson as exchanged particles. It should be noted that such type of approach has been used in the pioneering works~\cite{Voloshin:1976ap,Tornqvist:1991ks} but suffer from a systematic expansion as offered by effective field theory approaches, where
the short-ranged interaction is given in terms of contact terms with adjustable low-energy constants. The characterization of short-range interaction provides critical insights into the formation mechanisms of the hadronic molecular states\cite{Zou:2025sae}. The vector-meson exchange considered here is one specific representation of the short-range dynamics. In such an OBE model, the potential may contain a $\delta(\bm r)$ term representing short-range interactions. Two strategies exist in the literature for handling this $\delta(\bm r)$ term: either retaining it~\cite{0808.0073,Wang:2020dya,Chen:2021tip,Wang:2022mxy,Wang:2024ukc} or discarding it~\cite{Thomas:2008ja,Liu:2019zvb,Ling:2021asz}. However, since the short-range interaction between hadrons cannot be definitively determined by such a phenomenological model and may receive contributions from heavier particle exchanges, we introduce a parameter $a$ to phenomenologically adjust the strength of the $\delta(\bm r)$ term. This parameter effectively introduces an additional contact interaction to account for extra short-range interactions from other heavier meson exchanges. As demonstrated in our previous study \cite{Yalikun:2021bfm}, a specific value of parameter $a$ enables the interpretation of the four observed $P_c$ states with a simultaneous cutoff. In this work, we systematically investigate the analogous mechanism associated with the $\delta(\bm r)$ term within the $D^{(*)}\bar{D}^{(*)}$ and $B^{(*)}\bar{B}^{*}$ systems.

This paper is organized as follows. The details of the OBE model in the $D^{(*)}\bar D^{(*)}$ and the $B^{(*)}\bar B^{*} $ system as well as the terminology of the scattering matrix from the stationary Schr\"odinger equation are introduced in Sec.~\ref{sec:2}. Numerical results and discussions of the possible molecular states in the $D^{(*)}\bar D^{(*)}$ and $B^{(*)}\bar B^{*} $ systems are given in Sec.~\ref{sec:res}. Finally,  our conclusion are presented in Sec.~\ref{sec:summary}. 

\section{Effective Lagrangian and Potential}\label{sec:2}
The OBE potential model is quite successful in interpreting the formation mechanism of pentaquarks~\cite{He:2019rva,Chen:2019asm,Liu:2019zvb,Du:2021fmf,Yalikun:2021bfm}. In this work, we systematically study the OBE potentials in $D^{(*)}\bar D^{(*)}/B^{(*)}\bar B^{(*)} $ systems, and investigate the possibility of the hidden-charm or bottom tetraquarks states in the molecular picture.

To investigate the coupling between a charmed or bottomed meson with light scalar, pseudoscalar and vector mesons, we employ the effective Lagrangian satisfying chiral symmetry and heavy quark spin symmetry (HQSS), developed in Refs.~\cite{Cheng:1992xi,Yan:1992gz,Wise:1992hn,Cho:1994vg,Casalbuoni:1996pg,Pirjol:1997nh,Liu:2011xc},
\begin{align}\label{lag}
		\mathcal{L}_{H}&=g_S{\rm Tr}[ \bar H_a^{\bar Q} \sigma H_a^{\bar Q}]+
		ig {\rm Tr}[ \bar H_a^{\bar Q} \gamma\cdot A_{ab}\gamma^5H_b^{\bar Q}]\notag\\
		&-i\beta{\rm Tr}[\bar H_a^{\bar Q} v_\mu(\Gamma^\mu_{ab}-\rho^\mu_{ab})H_b^{\bar Q}]+i\lambda{\rm Tr}\left[ \bar H_a^{\bar Q} \frac{i}{2}[\gamma_\mu,\gamma_\nu]F^{\mu\nu}_{ab}H_b^{\bar Q}\right]\nonumber\\
		&+g_S{\rm Tr}[ H_a^{Q} \sigma \bar H_a^{ Q}]+ig {\rm Tr}[ H_a^{ Q} \gamma\cdot A_{ab}\gamma^5\bar H_b^{ Q}]\notag\\
		&+i\beta{\rm Tr}[H_a^{ Q} v_\mu(\Gamma^\mu_{ab}-\rho^\mu_{ab})\bar H_b^{ Q}]+i\lambda{\rm Tr}\left[ H_a^{ Q} \frac{i}{2}[\gamma_\mu,\gamma_\nu]F^{\mu\nu}_{ab}\bar H_b^{ Q}\right],
	\end{align}
with $a,b$ and $c$ the flavor indices and $v^\mu$ the four-velocity of the heavy hadron. The axial-vector and vector currents read, respectively,
\begin{align}
A^\mu &=\frac{1}{2}(\xi^\dagger\partial^\mu\xi-\xi\partial^\mu\xi^\dagger)=\frac{i}{f_\pi}\partial^\mu\mathbb{P}+\cdots, \notag\\
\Gamma^\mu &=\frac{i}{2}(\xi^\dagger\partial^\mu\xi+\xi\partial^\mu\xi^\dagger)=\frac{i}{2f_\pi^2}[\mathbb{P},\partial^\mu\mathbb{P}]+\cdots, 
\end{align}
with $\xi={\rm exp}(i\,\mathbb{P}/f_\pi)$ and $f_\pi=132$~MeV the pion decay constant. The vector meson fields $\rho^\alpha$ and field strength tensor $F^{\alpha\beta}$ are defined as $\rho^\alpha={i\,g_V}\mathbb{V}^\alpha/{\sqrt 2}$ and $F^{\alpha\beta}=\partial^\alpha\rho^\beta-\partial^\beta\rho^\alpha+[\rho^\alpha,\rho^\beta]$ with
\begin{align}
\mathbb{P}&=
\begin{pmatrix}
\frac{\pi^0}{\sqrt 2}+\frac{\eta}{\sqrt 6}&\pi^+&K^+\\
\pi^-&-\frac{\pi^0}{\sqrt 2}+\frac{\eta}{\sqrt 6}&K^0\\
K^-&\bar K^0&-\sqrt{\frac{2}{3}}\eta
\end{pmatrix},\\
\mathbb{V}&=
\begin{pmatrix}
\frac{\rho^0}{\sqrt 2}+\frac{\omega}{\sqrt 2}&\rho^+&K^{*+}\\
\rho^-&-\frac{\rho^0}{\sqrt 2}+\frac{\omega}{\sqrt 2}&K^{*0}\\
K^{*-}&\bar K^{*0}&\phi
\end{pmatrix},
\end{align}
where we have ignored the mixing between the pseudoscalar octet and singlet. Because the pseudoscalar singlet $\eta'$ has a much higher mass than the octet member $\eta$,  its contribution is highly suppressed in  low-energy systems as considered here. On the other hand, the $\eta'$-nucleon coupling is suppressed by a factor of about $1/3$ compared to the $\eta$ coupling in chiral perturbation theory~\cite{Feldmann:1998vh,Rosenzweig:1981cu}, a suppression pattern that is expected to persist in heavy meson systems. This octet approximation is frequently used in OBE models for heavy hadronic molecules~\cite{0808.0073,Thomas:2008ja,Wang:2020dya,Chen:2021tip,Wang:2022mxy,Wang:2024ukc,Liu:2019zvb,Ling:2021asz,Guo:2017jvc,Albaladejo:2017blx}.
The $S$-wave heavy meson can be represented by $H_a^{Q}$ and $H_a^{\bar Q}$, respectively, 
\begin{align}
	H_a^{\bar Q}&=(\pbfield^{\bar Q*}_{a\mu}\gamma^\mu-\pbfield^{\bar Q}_a\gamma^5)\frac{1-\slashed v}{2},\ \ 
	\bar H_a^{\bar Q}=\gamma^0H_a^{\bar Q\dagger}\gamma^0,\\
	H_a^{Q}&=\frac{1+\slashed v}{2}(\pfield^{Q*}_{a,\mu}\gamma^\mu-\pfield^Q_a\gamma^5),\ \ \bar H_a^{ Q}=\gamma^0H_a^{ Q\dagger}\gamma^0,
\end{align}
where heavy mesons with $J^P=0^-$ and $1^-$ are denoted by $\mathcal{P}$ and $\mathcal{P}^*_\mu$, respectively, which
are normalized as~\cite{Wise:1992hn,Wang:2020dya}
\begin{align}
    \langle 0 |\pbfield| \bar Q  q(0^-)\rangle=\sqrt{M_{\pbfield}}&,\quad    \langle 0|\pbfield^*_\mu|\bar Q  q(1^-)\rangle=\epsilon_\mu\sqrt{M_{\pbfield^*}},\\
	\langle 0 |\mathcal{P}| Q \bar q(0^-)\rangle=\sqrt{M_{\mathcal{P}}}&,\quad    \langle 0|\mathcal{P}^*_\mu|Q \bar q(1^-)\rangle=\epsilon_\mu\sqrt{M_{\mathcal{P}^*}},
\end{align} 
where $M_{\pfield^{(*)}}$ and $M_{\pbfield^{(*)}}$ are the masses 
of the heavy and anti-heavy mesons, and the polarization vector for heavy vector mesons is taken as $\varepsilon^\mu=(1,\bm\epsilon)$ in the heavy quark mass limit.  

Using the Lagrangian in Eq.~\eqref{lag}, we can derive the potentials from the OBE for the $\pfield^{(*)} \pbfield^{(*)}$ systems in the Breit approximation. The potential in momentum space reads
\begin{align}
\mathcal{V}^{h_1h_2\to h_3h_4}(\bm q)=-\frac{\mathcal{M}^{h_1h_2\to h_3h_4}}{\sqrt{2m_12m_22m_32m_4}},\label{eq:Breit}
\end{align}
 where $m_i$ is the mass of the particle $h_i$, $\bm q$ is the three momentum of the exchanged meson and
 $\mathcal{M}^{h_1h_2\to h_3h_4}$ is the scattering amplitude of the transition $h_1h_2\to h_3 h_4$. The Feynman diagrams for $\pfield^{(*)} \pbfield^{(*)}\to \pfield^{(*)} \pbfield^{(*)}$ are shown in Fig.\ref{fig:fynman-diag}. The momentum space potentials are collected in Appendix~\ref{sec:potential}.  
 \begin{figure}[ht]
	\centering
	\includegraphics[width=0.45\textwidth]{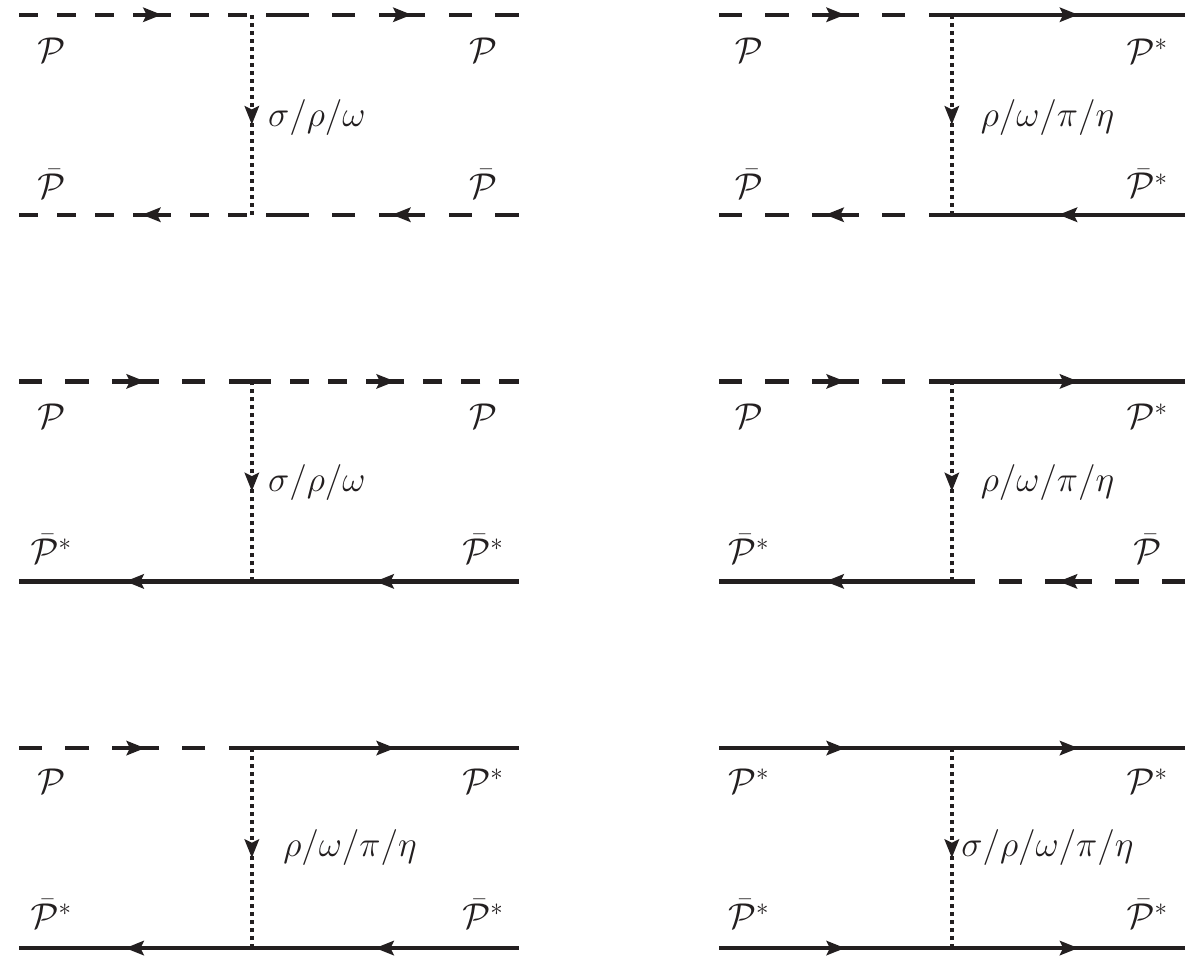}
	\caption{Feynman diagrams for $\pfield^{(*)} \pbfield^{(*)}\to\pfield^{(*)} \pbfield^{(*)}$ transition.}\label{fig:fynman-diag}
\end{figure}
 
The potentials in position space are obtained by performing the Fourier transformation,
\begin{align}
\mathcal{V}(\bm r,\Lambda,\mu_{\rm{ex}})=\int\frac{d^3\bm q}{(2\pi)^3} \mathcal{V}(\bm q)F^2(\bm q,\Lambda,\mu_{\rm{ex}}){\rm{e}}^{i\bm q \cdot \bm r},\label{eq:Fourier-trans}
\end{align} 
where the form factor with the cutoff $\Lambda$ is introduced to account for the inner structures of the interacting hadrons~\cite{Tornqvist:1993ng},
\begin{align}
F(\bm q,\Lambda,\mu_{\rm{ex}})=\frac{m_{\rm{ex}}^2-\Lambda^2}{(q^0)^2-\bm q^2-\Lambda^2}=\frac{\tilde{\Lambda}^2-\mu_{\rm{ex}}^2}{\bm q^2+\tilde{\Lambda}^2}.\label{eq:form-factor}
\end{align}
We notice that the form factor may break the symmetries we employed in the Lagrangian~\eqref{lag}. However, in the near-threshold region, this effect is expected to be not significant.
We have defined $\tilde\Lambda=\sqrt{\Lambda^2-(q^0)^2}$ and $\mu_{\rm{ex}}=\sqrt{m_{\rm{ex}}^2-(q^0)^2}$ for convenience. Note that, for inelastic scattering, the energy of the exchanged meson is nonzero, so the denominator of the propagator can be rewritten as $q^2-m_{\rm{ex}}^2=(q^0)^2-\bm q^2 - m_{\rm{ex}}^2=-(\bm q^2+\mu_{\rm{ex}}^2)$, with $\mu_{\rm{ex}}$ the effective mass of the exchanged meson. 
We note that the potentials involve three types of functions, $1/(\bm q^2+\mu^2_{\rm {ex}})$, $\bm A\cdot \bm q \bm B\cdot \bm q/(\bm q^2+\mu^2_{\rm {ex}})$ and $(\bm A\times \bm q) \cdot(\bm B\times \bm q)/(\bm q^2+\mu^2_{\rm {ex}})$, where $\bm A$ and $\bm B$ refer to the vector operators acting on the spin-orbit wave functions of the initial or final states, and their specific forms can be deduced from the corresponding terms in Eqs.~(\ref{eq:poten-in-p}). Therefore, to obtain the position space potentials, it is sufficient to perform the Fourier transformation on these three types of functions. The Fourier transformation of $1/(\bm q^2+\mu^2_{\rm {ex}})$, denoted by $Y_{\rm{ex}}$, reads
\begin{align}
Y_{\rm{ex}}&=\int\frac{d^3\bm q}{(2\pi)^3} \frac{1}{\bm q^2+\mu^2_{\rm{ex}}} \left (\frac{\tilde{\Lambda}^2-\mu^2_{\rm{ex}}}{\bm q^2+\tilde{\Lambda}^2}\right )^2 e^{i\bm q\cdot \bm r},\notag\\
&=\frac{1}{4\pi r}({\rm{e}}^{-\mu_{\rm{ex}}r}-{\rm{e}}^{-\tilde{\Lambda}r})-\frac{\tilde{\Lambda}^2-\mu_{\rm{ex}}^2}{8\pi\tilde{\Lambda}}{\rm{e}}^{-\tilde{\Lambda}r}.\label{eq:Yex}
\end{align}     
Before performing the Fourier transformation on $\bm A\cdot \bm q \bm B\cdot \bm q/(\bm q^2+\mu_{\rm{ex}}^2)$, we can decompose it as 
\begin{align}
\frac{\bm A\cdot \bm q \bm B\cdot \bm q}{\bm q^2+\mu_{\rm{ex}}^2}&=\frac{1}{3}\left \{\bm A\cdot\bm B\left (1-\frac{\mu_{\rm{ex}}^2}{\bm q^2+\mu_{\rm{ex}}^2}\right )-\frac{S(\bm A,\bm B,\hat q)|\bm q|^2}{\bm q^2+\mu_{\rm{ex}}^2}\right \}, \label{eq:CT-pot}
\end{align}
where $S(\bm A,\bm B,\hat q)=3 \bm A\cdot \hat q \bm B\cdot \hat q-\bm A\cdot\bm B$ is the tensor operator in momentum space. It can be found that without the form factor, the constant term in Eq.~\eqref{eq:CT-pot} leads to a $\delta(\bm r)$ term in coordinate space after the Fourier transformation. With the form factor, the $\delta(\bm r)$ term becomes finite, and it dominates the short-range part of the potential. In the phenomenological view, the $\delta(\bm r)$ term can mimic the role of contact interaction~\cite{Yalikun:2021bfm}, which is also related to the regularization scheme~\cite{Tornqvist:1993ng}. In Refs.~\cite{Thomas:2008ja,Yamaguchi:2017zmn}, after removing the $\delta(\bm r)$ term, the hadronic molecular picture for some observed hidden-charm states is discussed with the one-pion-exchange potential, which is assumed to be of long-range. In this work, we will separately analyze the poles in the system with or without the $\delta(\bm r)$ term. For this purpose, we introduce a parameter $a$ to distinguish these two case, 
\begin{align}
\frac{\bm A\cdot \bm q \bm B\cdot \bm q}{\bm q^2+\mu_{\rm{ex}}^2}-\frac{a}{3}\bm A\cdot \bm B&=\frac{1}{3}\left \{\bm A\cdot\bm B\left (1-a-\frac{\mu_{\rm{ex}}^2}{\bm q^2+\mu_{\rm{ex}}^2}\right )\right.\notag\\
&\ \ \ \ \left.-S(\bm A,\bm B,\hat q)\frac{|\bm q|^2}{\bm q^2+\mu_{\rm{ex}}^2}\right \}. \label{eq:CT-pot1}
\end{align}
After performing the Fourier transformation of Eq.~\eqref{eq:CT-pot1}, we have
\begin{align}
&\int\frac{d^3\bm q}{(2\pi)^3} \left (\frac{\bm A\cdot \bm q \bm B\cdot \bm q}{\bm q^2+\mu_{\rm{ex}}^2}-\frac{a}{3}\bm A\cdot \bm B\right ) \left (\frac{\tilde{\Lambda}^2-\mu^2_{\rm{ex}}}{\bm q^2+\tilde{\Lambda}^2}\right )^2 e^{i\bm q\cdot \bm r}\notag\\
&=-\frac{1}{3}[\bm A\cdot \bm BC_{\rm{ex}}+S(\bm A,\bm B,\hat r)T_{\rm{ex}}]\label{eq:CT-pot2},
\end{align}
where $S(\bm A,\bm B,\hat r)=3 \bm A\cdot \hat r \bm B\cdot \hat r-\bm A\cdot\bm B$ is the tensor operator in coordinate space, and the functions $C_{\rm{ex}}$ and $T_{\rm{ex}}$ read 
\begin{align}
C_{\rm{ex}}&=\frac{1}{r^2}\frac{\partial}{\partial r}r^2\frac{\partial}{\partial r}Y_{\rm{ex}}+\frac{a}{(2\pi)^3}\int \left (\frac{\tilde{\Lambda}^2-\mu^2_{\rm{ex}}}{\bm q^2+\tilde{\Lambda}^2}\right )^2 e^{i\bm q\cdot \bm r}d^3\bm q,\\
T_{\rm{ex}}&=r\frac{\partial}{\partial r}\frac{1}{r}\frac{\partial}{\partial r}Y_{\rm{ex}}.
 \end{align}   
Apparently, the contribution of the $\delta(\bm r)$ term is fully included (excluded) when $a=0(1)$~\cite{Yalikun:2021bfm,Wang:2020dya}. Similarly, the Fourier transformation of the function $(\bm A\times \bm q) \cdot(\bm B\times \bm q)/(\bm q^2+\mu^2_{\rm {ex}})$ can be evaluated with the help of the relation $(\bm A\times \bm q) \cdot(\bm B\times \bm q)=\bm A\cdot \bm B|\bm q|^2-\bm A\cdot \bm q \bm B\cdot \bm q$. 

With the prescription above, the coordinate space representations of the potentials in Eqs.~\eqref{eq:poten-in-p} can be written in terms of $Y_{\rm{ex}}$, $C_{\rm{ex}}$ and $T_{\rm{ex}}$ given in Eqs.~\eqref{eq:Yex} and ~\eqref{eq:CT-pot2}. The potentials should be projected into certain partial waves by sandwiching the spin operators in the potentials between the partial waves of the initial and final states. We refer to Refs.~\cite{Yalikun:2021bfm,Yalikun:2021dpk} for computing the partial wave projections. In this work, we focus on the positive parity states which are possibly bound in S-wave and more easily form the molecular states respect to negative ones. The partial waves corresponding to the spin-parities of $J^P=0^{+},1^+,2^+$ are shown in Table~\ref{tab:pw}.
\begin{table}[t]\centering
	\caption{Partial waves in the given $J^{P}$.}\label{tab:pw}
	\begin{ruledtabular}
	\begin{tabular}{cccc}
		&$\pfield\pbfield$&$\pfield\pbfield^*/\pfield^*\pbfield$&$\pfield^*\pbfield^*$\\\hline
		$J^p=0^{+}$&$^1S_{0}$&-&$^1S_0$,$^5D_0$\\
		$J^p=1^+$&-&$^3S_1$,$^3D_1$&$^3S_1$,$^3D_1$,$^5D_1$\\
		$J^p=2^+$&$^1D_2$&$^3D_2$&$^5S_2$,$^1D_2$,$^3D_2$,$^5D_2$\\
	\end{tabular}
\end{ruledtabular}
\end{table}

In our numerical calculation, the masses of exchanged particles are taken as $m_{\sigma}=600.0$ MeV, $m_{\pi}=138.0$ MeV, $m_{\eta}=547.9$ MeV, $m_{\rho}=770.7$ MeV and $m_{\omega}=782.0$ MeV. 
The $\sigma$ meson in our work refers to the lightest scalar meson with isospin $0$ and spin-parity $0^+$, corresponding to $f_0(500)$ in the RPP, which is a very broad state with a large mass uncertainty ($400-800$ MeV). It has been shown that contributions from such a broad resonance exchange (effectively correlated scalar-isoscalar $2\pi$ exchange) in $t$-channel can be represented by a stable particle with a mass about 600 MeV in nucleon-nucleon interaction~\cite{Meissner:1990kz,Wu:2023uva}. We simply set the mass of $\sigma$ to $600$ MeV, which is also commonly used in the OBE model for hadronic molecules~\cite{Liu:2009qhy,Liu:2019zvb,Xu:2025mhc}. The coupling constants in the Lagrangian can be extracted from experimental data or deduced from various theoretical models. Here we adopt the values given in Refs.~\cite{Ding:2008gr,Liu:2011xc,Meng:2019ilv,Isola:2003fh}, i.e.,  $g_S=0.76$, $g=-0.59$, $\beta=0.9$, $\lambda=0.56~\rm{~GeV}^{-1}$ and $g_V=5.9$, while their relative phases are fixed by the quark model~\cite{Riska:2000gd,Yalikun:2021dpk}.  
 With the coupled-channel potential matrix $\mathcal{V}_{jk}$, the radial Schr\"odinger equation can be written as  
\begin{eqnarray}
\left[-\frac{1}{2\mu_j}\frac{d^2}{d r^2}+\frac{l_j(l_j+1)}{2\mu_jr^2} + W_j\right]u_j+
\sum_{k}
\mathcal{V}_{jk}u_k=
E u_j,
\label{eq_schro_coupl}
\end{eqnarray}
where $j$ is the channel index; $u_j$ is defined by $u_j(r)=rR_j(r)$ with the radial wave function $R_j(r)$ for the $j$-th channel; $\mu_j$ and $W_j$ are the corresponding reduced mass and threshold; $E$ is the total energy of the system. The momentum for channel $j$ is expressed as 
\begin{align}\label{eq:ch-mom}
q_j(E)=\sqrt{2\mu_j(E-W_j)}.
\end{align}
 By solving Eq.~\eqref{eq_schro_coupl}, we obtain the wave function which is normalized to satisfy the incoming boundary condition for the $j$-th channel~\cite{osti_4661960},
\begin{eqnarray}
u_{j}^{(k)}(r)\overset{r\rightarrow \infty}{\longrightarrow} \delta_{jk}e^{-iq_j r}-S_{jk}(E)e^{iq_j r}\label{eq:asym-wave},
\end{eqnarray}
where $S_{jk}(E)$ is the scattering matrix component. In the multi-channel problem, there is a sequence of thresholds, $W_1<W_2<\cdots$, and the scattering matrix element $S_{jk}(E)$ is an analytic function of $E$ except at the branch points $W_j$ and possible poles. Bound/virtual states and resonances are represented as the poles of the $S_{jk}(E)$ on the complex energy plane~\cite{osti_4661960}. 

The characterization of these poles requires analytical continuation of the $S$ matrix to the complex energy plane, where the poles must be searched for on the correct Riemann sheet (RS). Since the momentum $q_j$ is a double-valued function of energy $E$, each channel has two RSs: the first (physical) sheet where ${\rm{Im}}[q_j]\geq 0$, and the second (unphysical) sheet where ${\rm{Im}}[q_j]< 0$. In an $n$-channel system, the scattering amplitude has $2^n$ RSs, each labeled by $(\pm,\cdots,\pm)$ where the $j$-th ``$\pm$" indicates the sign of ${\rm{Im}}[q_j(E)]$. Note that the form factor introduced in Eq.~\eqref{eq:form-factor} does not introduce additional non-analytic structures into the scattering amplitude within the near-threshold energy region we consider.

\section{Results and Discussion}\label{sec:res}
\subsection{$D^{(*)}\bar D^{(*)}$ systems}
In this subsection, we discuss the near-threshold molecular states in $D^{(*)}\bar D^{(*)}$ systems, which are not only important to understand the molecular nature of hidden-charm tetraquarks, but also a good starting point to extend our analysis to other hidden-heavy quark systems. We use the isospin-averaged masses for the charmed mesons in the following calculations. $D^{(*)}\bar D^{(*)}$ can be grouped into four systems as $D\bar D$, $D\bar D^*$, $D^*\bar D$ and $D^*\bar D^*$. Among them, $ D \bar D^* $ and $D^* \bar D $ have the same mass and quark configuration, and they  mix~\cite{Liu:2008fh}   
\begin{align}
	|D \bar D^*\rangle_\pm= \frac{1}{\sqrt 2}[|D \bar D^*\rangle \pm |D^* \bar D\rangle],\label{eq:DD_charge}  
\end{align}
which represent the two charge-conjugation eigenstates of the $D \bar D^*$ systems, $\mathcal{C}|D \bar D^*\rangle_\pm=\mp|D \bar D^*\rangle_\pm$, with $\mathcal{C}$ the charge conjugation operator. 
The OBE potentials are expressed as 
\begin{align}
	\mathcal{V}^{D^{(*)}\bar D^{(*)}\to D^{(*)}\bar D^{(*)}}= \sum\limits_{\rm{ex}=\sigma,\pi,\eta,\rho,\omega}I_{\rm{ex}}\mathcal{V}^{\pfield^{(*)} \pbfield^{(*)}\to \pfield^{(*)} \pbfield^{(*)}}_{\rm{ex}}	
\end{align}
where $I_{\rm{ex}}$ stands for the isospin factors shown in Table~\ref{tab:iso}.  
\begin{table}[t]\centering
	\caption{Isospin factors for each meson exchange in $D^{(*)}\bar D^{(*)}\to D^{(*)}\bar D^{(*)}$ transition, where $I=0$ or $I=1$ representing the total isospin}\label{tab:iso}
	\begin{ruledtabular}
	\begin{tabular}{cccccc}
		&$\sigma$&$\pi$&$\eta$&$\rho$&$\omega$\\\hline
		$I_F(I=0)$ & $1$ & $3/2$ & $1/6$ & $3/2$ & $1/2$\\
		$I_F(I=1)$ & $1$ & $-1/2$ & $1/6$ & $-1/2$ & $1/2$\\
	\end{tabular}
\end{ruledtabular}
\end{table}

\begin{figure}[t]
	\centering
	\includegraphics[width=\linewidth]{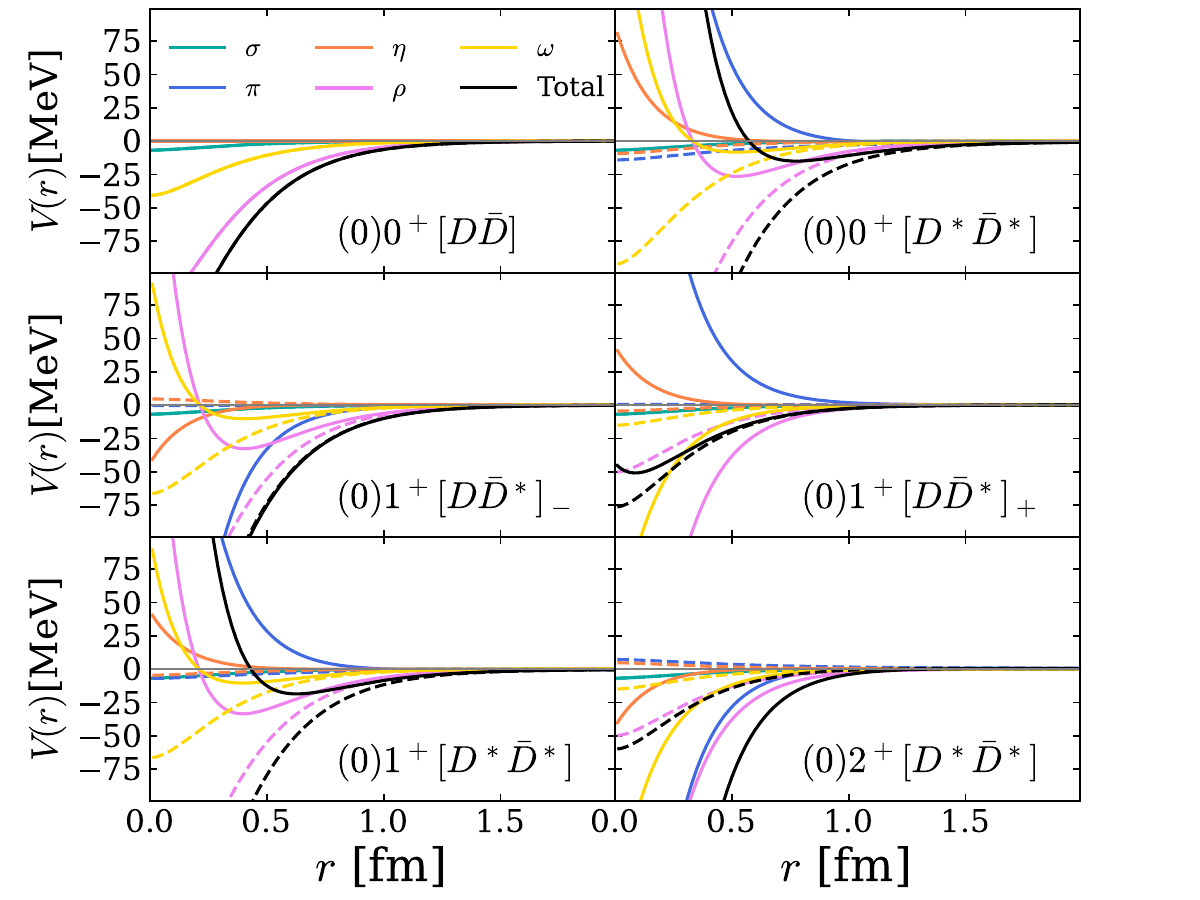}
	\caption{Single channel potentials in $S$-wave for $I=0$ systems with $\Lambda=1.2$~GeV. Quantum numbers are denoted by $(I)J^P$, and the dashed lines represents the case without $\delta(\bm r)$ term.}\label{fig:pot-I0}
\end{figure}

\begin{figure}[t]
	\centering
	\includegraphics[width=\linewidth]{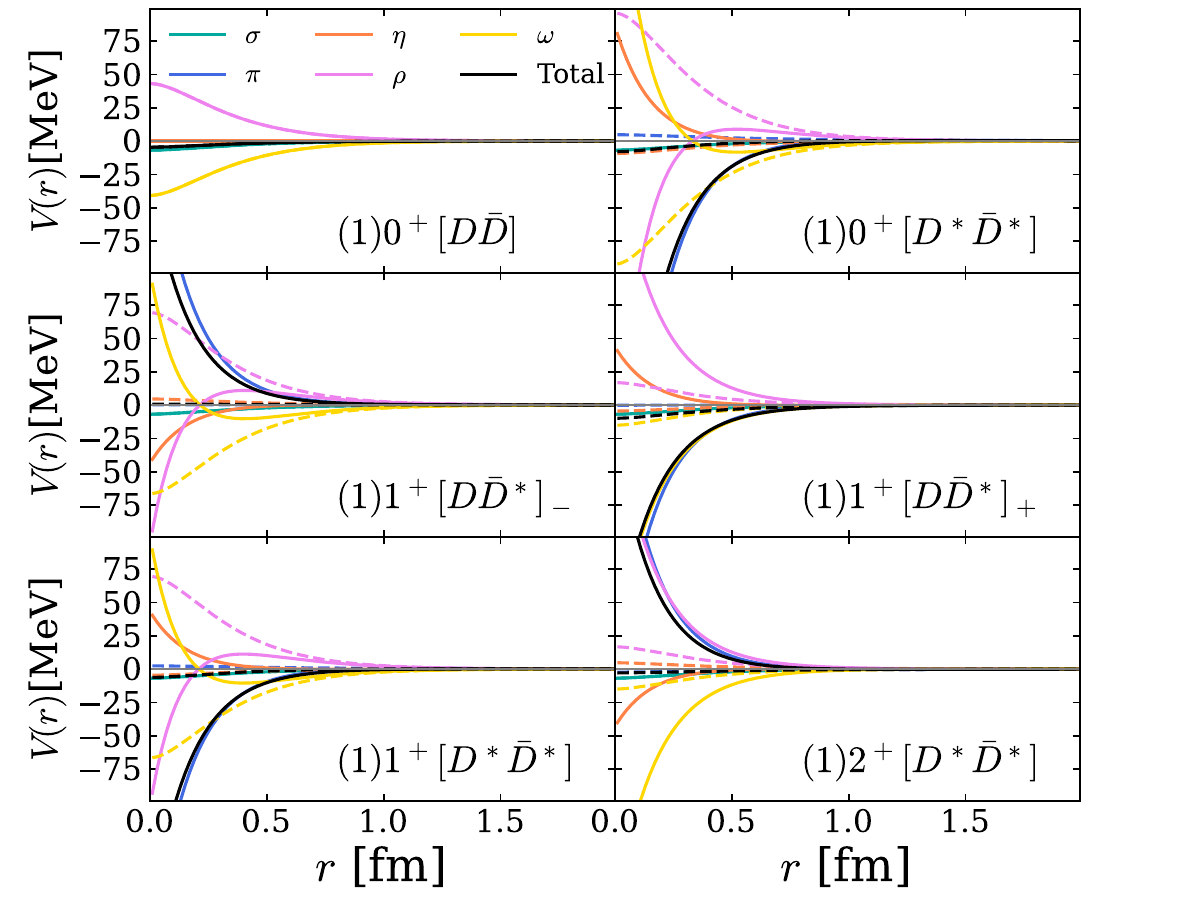}
	\caption{Single channel potentials in $S$-wave for $I=1$ systems with $\Lambda=1.2$~GeV. See the caption of Fig.~\ref{fig:pot-I0}.}\label{fig:pot-I1}
\end{figure}

The S-wave potentials for $I=0$ systems are shown in Fig.~\ref{fig:pot-I0}, while those for $I=1$ are shown Fig.~\ref{fig:pot-I1}, where the strength of individual meson exchange potentials in both cases, with- and without-$\delta(\bm r)$ terms, are compared. {The general conclusion is that the $I=0$ system is more attractive then the corresponding $I=1$ system, and the potentials of $[D \bar D^*]_\pm$ and $D^* \bar D^*$ systems with various spin-parity quantum numbers $J^P$ depends on the $\delta(\bm r)$ terms.}    

\begin{figure}[ht]
	\centering
	\subfigure[With $\delta$-term]{\label{sfig:BEd0}\includegraphics[width=0.9\linewidth]{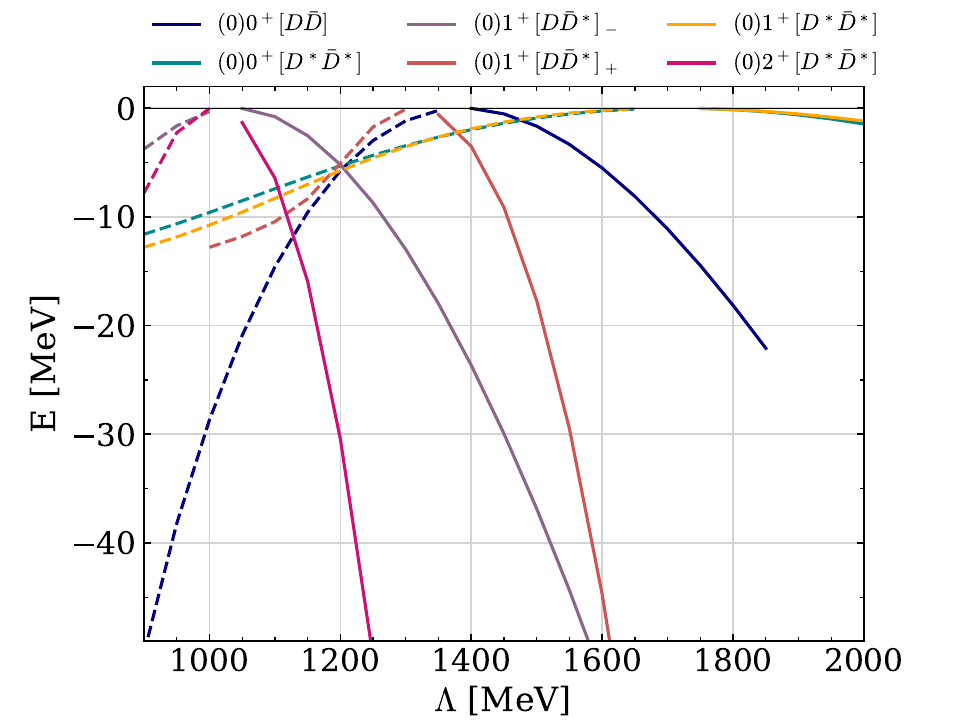}}
	\subfigure[Without $\delta$-term]{\label{sfig:BEd1}\includegraphics[width=0.9\linewidth]{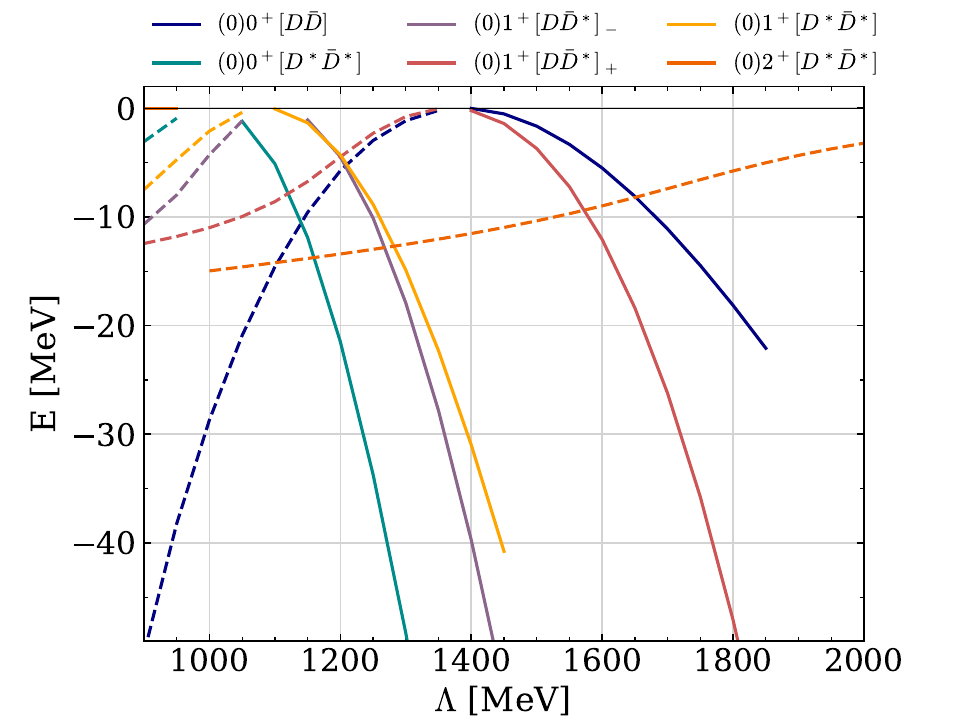}}
	\caption{Binding energy of the bound states (solid curves) or virtual states (dashed curves) in the single channels isoscalar systems as $\Lambda$ increases.}\label{fig:BE-I0}
\end{figure}

In the single channel case, we are dealing with either bound or virtual states, which appear as poles of the scattering amplitude on the real axis of the complex energy plane. The binding energy is defined as   
 \begin{align}
 \mathbb{B}=E_{\mathrm{pole}}-W.
 \end{align}
 As discussed in the previous section, the $\delta(\bm r)$ term dominates the short-range dynamics of the potentials, and thus it serves as the phenomenological contact term. It is seen that the proper treatment of the $\delta(\bm r)$ term in the OBE model plays an important role in the simultaneous interpretation of the $P_c$ states~\cite{Yalikun:2021bfm}. Therefore, we will represent the results in two extreme cases, with or without the $\delta(\bm r)$ term. The binding energies of the states with $I=0$ as cutoff varies from 1 to 2~GeV are shown in Fig.~\ref{fig:BE-I0}, where the $S$-$D$ wave mixing effects are included. The sub-figures~\ref{sfig:BEd1} and~\ref{sfig:BEd0} show the effects of the $\delta(\bm r)$ term on the binding energies of the single channel systems. It is found that the effects of the $\delta(\bm r)$ term are more significant in $D^*\bar D^*$ systems than in $[D\bar D^*]_\pm$ systems but the $D\bar D$ system is independent of it. 
 The interpretation can be elucidated by examining the OBE potentials associated with these channels, as depicted in Fig.~\ref{fig:pot-I0}. Specifically, the potential for the $D\bar{D}$ channel lacks any presence of the $\delta(\bm{r})$ term, while the potential for the $D^*\bar{D}^*$ channel receives significant modification upon the exclusion of the $\delta(\bm{r})$ term relative to the potentials of the $[D\bar{D}^*]_\pm$ channels. The latter observation results from the specific values of the coupling constants in the Lagrangian defined in Eq.~\eqref{lag}.
 
For the $I=0$ system, bound states can exist in all six channels when the cutoff parameter is sufficiently large. The $X(3872)$, with a PDG averaged mass of $3871.64\pm0.06$~MeV~\cite{ParticleDataGroup:2024cfk}, is understood as a molecular state below the $D\bar D^*$ threshold. Our analysis shows that the charge-conjugation eigenstate $(0)1^+[D\bar D^*]_-$ binds with $\Lambda\approx 1.2$~GeV, no matter the $\delta(\bm r)$ term is included or not. When the $\delta(\bm r)$ term is excluded (Fig.~\ref{sfig:BEd1}), the $(0)2^+D^*\bar D^*$ state becomes virtual, while $(0)0^{+}D^*\bar D^*$ and $(0)1^+D^*\bar D^*$ bind more easily compared to the case with $\delta(\bm r)$ (Fig.~\ref{sfig:BEd0}). The $\delta(\bm r)$ term has minimal impact on both $(0)1^+[D\bar D^*]_\pm$ states due to cancellation between vector and pseudoscalar meson exchange potentials (Fig.~\ref{fig:pot-I0}). For the $I=1$ system with the $\delta(\bm r)$ term, we find no bound states within the cutoff range $1\sim2$~GeV. However, virtual state poles appear in the $(1)0^{+}[D^*\bar D^*]$, $(1)1^+[D \bar D^*]_+$ and $(1)1^+[D^*\bar D^*]$ systems on the unphysical Riemann sheets, approaching their respective thresholds as the cutoff increases.

\begin{figure}[t]
	\centering
	\includegraphics[width=0.45\textwidth]{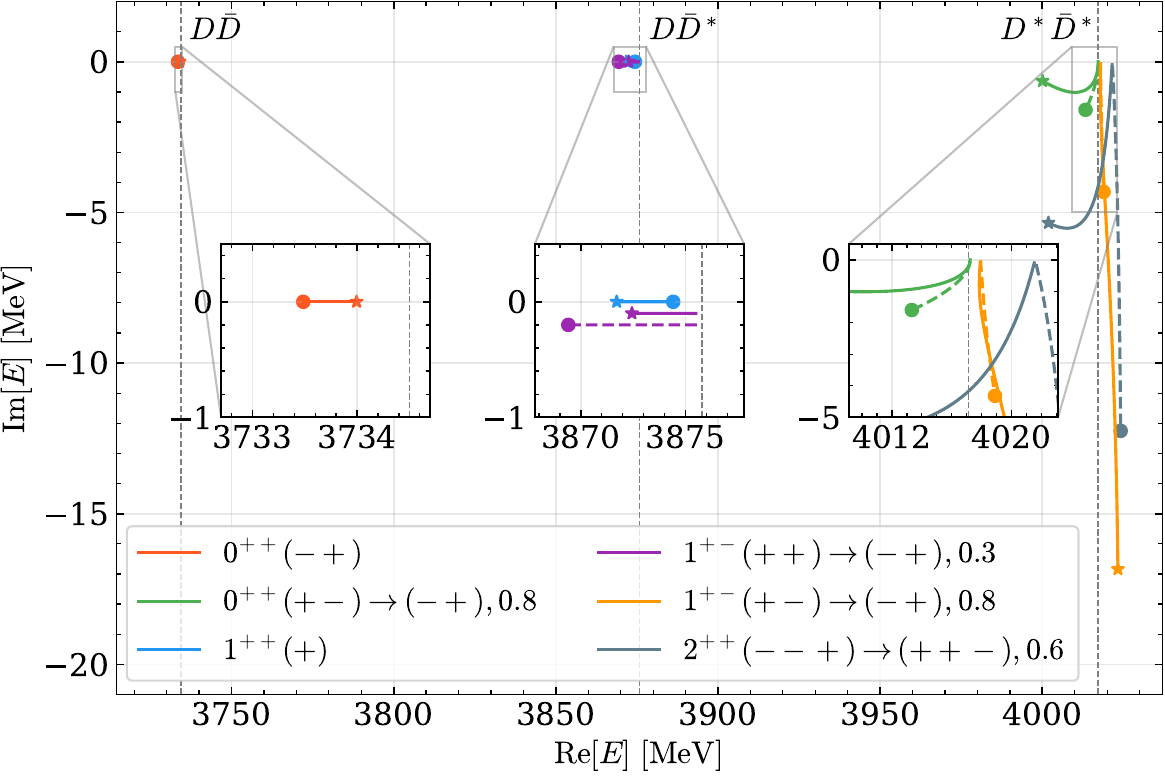}
    \caption{The trajectory of poles in the $D^{(*)}\bar{D}^{(*)}$ isoscalar system as the parameter $a$ varies from 0 to 1, with the cutoff $\Lambda$ set at 1.11 GeV. If a pole remains on the same Riemann sheet (RS) as $a$ varies from 0 to 1, its trajectory is shown as a solid line, with a star marking the starting point ($a=0$) and a circle marking the endpoint ($a=1$). When a pole transitions between different RS's, its trajectory is split into two curves: a solid line for $a=0$ to $a_0$ and a dashed line for $a=a_0$ to 1, where $a_0$ is the transition point specified in the legend.
    }
    \label{fig:pole_traject}
\end{figure}

Having analyzed the single-channel cases, we now turn our attention to the coupled-channel systems. Using the partial waves presented in Table~\ref{tab:pw}, we group channels with identical quantum numbers to construct several coupled-channel systems identified by their $J^{PC}$ as $0^{++}$, $1^{+\pm}$ and $2^{++}$. We explicitly include the $C$-parity to determine which channels can couple with each other due to $C$-parity conservation. Specifically, we study the coupled channel systems $D\bar D$-$D^*\bar D^*$, $[D\bar D^*]_+$-$D^*\bar D^*$ and $D\bar D$-$[D\bar D^*]_-$-$D^*\bar D^*$ for the quantum numbers $J^{PC}=0^{++}$, $1^{+-}$ and $2^{++}$, respectively. Additionally, we examine the $[D\bar D^*]_-$ channel alone for the $1^{++}$ system. For the isoscalar $1^{++}$ system, which corresponds to the quantum numbers of $X(3872)$, only the $[D\bar D^*]_-$ channel contributes, making it effectively a single-channel problem. Using the  measured mass of the  $X(3872)$ as input, we determine the cutoff parameter $\Lambda$ to be $1.11$ GeV. This value shows minimal dependence on the parameter $a$, as evidenced by the very short trajectory labeled by $1^{++}$ in Fig.~\ref{fig:pole_traject}. With this calibrated cutoff, we then calculate the trajectories of near-threshold poles in all $D^{(*)}\bar D^{(*)}$ systems as $a$ varies from 0 to 1.

In the $(0)0^{++}$ systems, we find two poles: a virtual pole located approximately 1 MeV below the $D\bar D$ threshold, and another pole near the $D^*\bar D^*$ threshold. If we consider a systematic uncertainty by varying $\Lambda$ slightly, the first pole can move to the physical RS, becoming a bound state. This state has been extensively discussed in the literature, with predictions from various phenomenological models~\cite{Zhang:2006ix,Gamermann:2006nm,Liu:2009qhy,Wong:2003xk,Nieves:2012tt,Hidalgo-Duque:2012rqv} and lattice QCD calculations~\cite{Prelovsek:2020eiw}. However, despite several studies~\cite{Gamermann:2007mu,Dai:2020yfu,Wang:2020elp} analyzing available experimental data~\cite{Belle:2005rte,Belle:2007woe,BaBar:2010jfn}, no clear experimental evidence has yet been found. It is worth noting that distinguishing between bound and virtual poles using only the $D\bar D$ distribution is challenging, as they can produce identical lineshapes above threshold~\cite{Guo:2017jvc}. The pole near the $D^*\bar D^*$ threshold shows a notable dependence on $a$, as illustrated by the green trajectory in Fig.~\ref{fig:pole_traject}. As $a$ increases from 0 to 1, this pole moves from RS($--$) to RS($-+$), indicating a transition from a virtual state of $D^*\bar D^*$ to a bound state.

For the $(0)1^{+-}$ systems, we find two near-threshold poles: one near the $D\bar D^*$ threshold as the $C$-parity partner of $X(3872)$, and the other near the $D^*\bar D^*$ threshold. The former evolves from a virtual state to a bound state as $a$ increases. Note that the trajectory of it lies on the real energy axis and is slightly shifted by hand in the plot for it is visibility.  This state, denoted as $\tilde{X}(3872)$ in the COMPASS collaboration's analysis~\cite{COMPASS:2017wql}, has been discussed in several theoretical works~\cite{Gamermann:2007fi,Wang:2020dgr,Dong:2021juy} as the $C$-parity partner of $X(3872)$ with mass degeneracy. The COMPASS analysis revealed preliminary evidence through distinctive kinematic features in the $J/\psi\pi^+\pi^-$ decay channel, particularly manifest in the two-pion invariant mass distribution that differs significantly from $X(3872)$'s characteristic pattern. Our theoretical work demonstrates that the classification of this state - whether as a bound or virtual state - exhibits critical dependence on the parameter $a$ governing short-range interaction dynamics in the OBE model.    

For the $(0)2^{++}$ system, $D\bar D$ and $D\bar D^*$ can not be in $S$-wave and hence no poles are found near their thresholds. Near the $D^*\bar D^*$ threshold, a pole moves from RS$(--+)$, corresponding to a $D^*\bar D^*$ bound state, to RS($++-$), turning into a virtual state. By imposing HQSS, it leads to the prediction of a $2^{++}$ $D^*\bar D^*$ tensor state as the HQSS partner of the $X(3872)$ considering the physical charmed meson masses~\cite{Nieves:2012tt,Hidalgo-Duque:2012rqv,Guo:2013sya}. It was argued in Ref.~\cite{Shi:2023ntq} that the state observed by Belle collaboration~\cite{Belle:2021nuv} is a good candidate of such a molecular state.

Similar to the coupled channel analysis of isoscalar systems above, now we study the isovector coupled channel systems with $J^{PC}=0^{++},1^{+\pm}$ and $2^{++}$. However, no poles (virtual or resonance) are found within the cutoff range of $1\sim2$~GeV, which indicates that any such poles would be located far from the physical real energy axis. This is consistent with our single channel analysis where the OBE potentials for $I=1$ cases were found to be insufficiently attractive to form bound or virtual states. The absence of isovector poles in our work is consistent  with fundamental the OBE dynamics as the isovector channels exhibits reduced attraction due to the cancellations between isovector ($\pi,\rho$) and isoscalar ($\eta,\omega$) meson exchanges (see Fig.~\ref{fig:pot-I1}). This contrasts with experimental observations of isovector states in the $Z_c$ family, suggesting either a sizable compact tetraquark component in such isovector states~\cite{Esposito:2016noz} or molecular configurations with additional short-range forces beyond meson-exchange~\cite{Chen:2022asf}.

\subsection{$B^{(*)}\bar B^{(*)}$ systems}

\begin{figure}[t]
	\centering
	\includegraphics[width=0.45\textwidth]{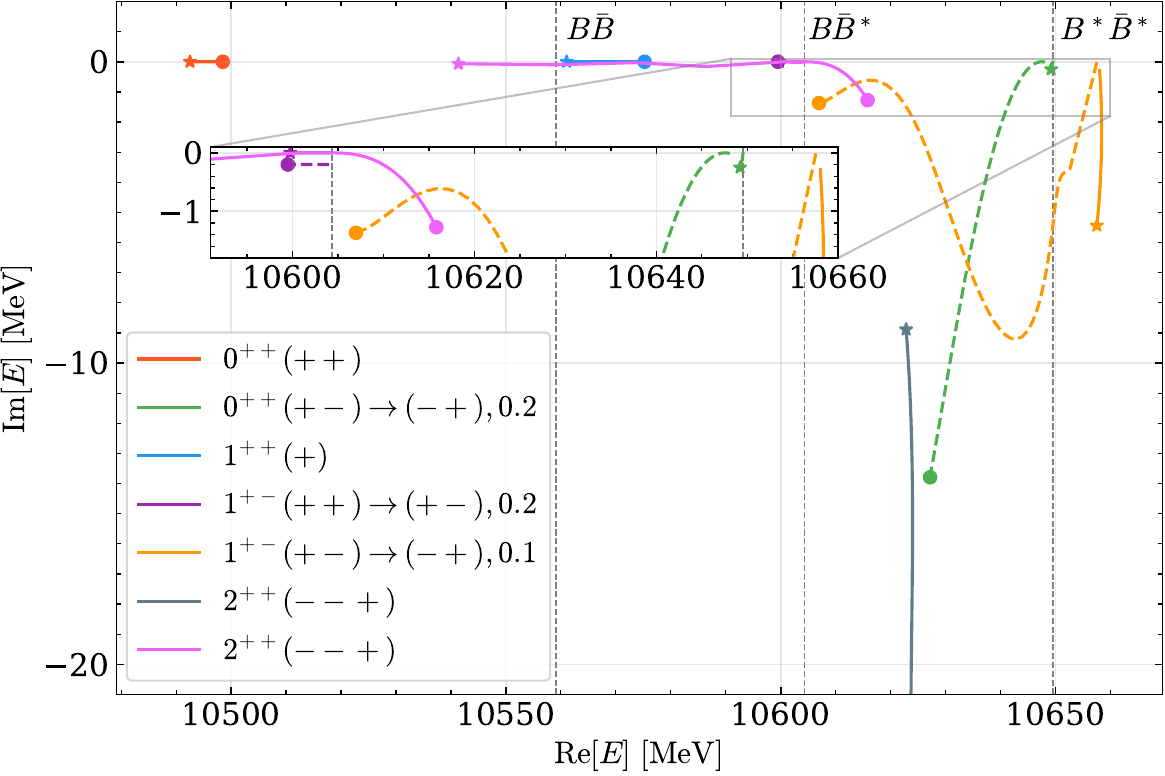}
    \caption{The trajectory of poles in the $B^{(*)}\bar{B}^{(*)}$ isoscalar system as the parameter $a$ varies from 0 to 1, with the cutoff $\Lambda$ set at 1.11 GeV. See the caption of Fig.~\ref{fig:pole_traject}.
    }
    \label{fig:pole_traject_B}
\end{figure}

The OBE potentials for $B^{(*)}\bar B^{(*)}$ systems are similar to those for $D^{(*)}\bar D^{(*)}$ systems. Therefore, we move directly to the coupled-channel analysis and skip the calculation of single-channel bound states or virtual states. For the $B^{(*)}\bar B^{(*)}$ systems, coupled-channel dynamics of the $B\bar B$, $[B\bar B^*]_\pm$ and $B^*\bar B^*$ channels are considered. We use the same strategy as in the hidden charm sector: $\Lambda$ is fixed to be $1.11$~GeV and $a$ is varied from $0$ to $1$ to investigate the effects of the $\delta(\bm r)$ term. The pole trajectories are shown in Fig.~\ref{fig:pole_traject_B}.              

In the $B\bar B$-$B^*\bar B^*$ system with $(0)0^{++}$, a bound state pole is found and its mass is below the $B\bar B$ threshold by about 60~MeV. The $\delta(\bm r)$ term contributes a slightly attractive force to this bound state, which originated in the inelastic potential described by $\pfield\pbfield\to\pfield^*\pbfield^*$ transition. A narrow resonance below the $B^*\bar B^*$ threshold is also found in this coupled channel system when $a=0$. By increasing the contribution of the $\delta(\bm r)$ term, this resonance is pushed toward the physical region on the RS $(-+)$ and then moves to RS($+-$), becoming a virtual state due to its repulsive contribution to the single channel potential $\pfield^*\pbfield^*$, which can be deduced from the S-wave $D^*\bar D^*$ potential in Fig.~\ref{fig:pot-I0}.

For the $(0)1^{++}$ channel, we find a bound state at about 30 MeV below the $B\bar B^*$ threshold. This is the analogue of $X(3872)$ in the hidden bottom sector, denoted as $X_b$. Although predicted in many studies, see e.g. Refs.~\cite{Guo:2013sya,Karliner:2014lta}, only negative experimental results~\cite{CMS:2013ygz,ATLAS:2014mka} have been reported so far. The reason may be that the decay $X_b\to \Upsilon(nS)\pi^+\pi^-$ violates isospin symmetry and $\Upsilon(nS)\pi^+\pi^-\pi^0$ is a better channel to search for $X_b$~\cite{Guo:2013sya}.

For the $(0)1^{+-}$ system, we found two near-threshold poles, one analogous to $\tilde X(3872)$ near the $B\bar B^*$ threshold and the other near the $B^*\bar B^*$ threshold, both being either a bound or virtual state depending on the value of $a$.

In the $(0)2^{++}$ system, two poles are found on the first RS $(--+)$ with $a=0$. The first pole lies below the $B\bar{B}$ threshold with a tiny imaginary part, while the second pole resides below the $B^*\bar{B}^*$ threshold and exhibits a relatively larger imaginary part. As $a$ increases, the former pole moves toward the $B^*\bar{B}^*$ threshold and takes a position on the complex energy plane connected to the physical energy axis when $a$ is close to 1, which is significant and causes a peak-like structure in the amplitude. However, the latter pole gradually moves away from the physical energy axis and does not cause a visible impact on the amplitude.

The comparison between the $D^{(*)}\bar{D}^{(*)}$ and $B^{(*)}\bar{B}^{(*)}$ systems within the isoscalar sector reveals that both exhibit comparable characteristics in their hadronic molecular spectra. The $B^{(*)}\bar{B}^{(*)}$ systems demonstrate stronger attractions compared to the hidden charm systems, a phenomenon attributed to their lower kinetic energy resulting from the larger reduced mass of the constituent particles. It is further suggested that applying a uniform theoretical framework or cutoff for both systems leads to more deeply bound molecular states in the $B^{(*)}\bar{B}^{(*)}$ sector than in the $D^{(*)}\bar{D}^{(*)}$ sector. Our results suggest that several isoscalar bottomonium-like exotic tetraquark states near the $B^{(*)}\bar B^{(*)}$ thresholds exist and can be potentially discovered in future experiments. For the isovector $B^{(*)}\bar B^{(*)}$ systems, we still cannot find any poles in the coupled channel analysis of $J^{PC}=0^{++},1^{+\pm}$ and $2^{++}$ states near the thresholds.

In previous coupled-channel analyses, we investigated the dependence of various poles on the parameter $ a $ while keeping the cutoff fixed at 1.11 GeV. Here, we briefly examine the sensitivity of these poles to variations in the cutoff by considering a range from 1.06 to 1.16 GeV. The cutoff variation is set in this narrow region so that most poles move in the same RS and it is convenient to observe the sensitivity of these poles to the cutoff, which is already sufficient to see the trend of pole variations. Specifically, we focus on the pole positions in the isoscalar $ D^{(*)} \bar{D}^{(*)} $ and $ B^{(*)} \bar{B}^{(*)} $ systems with the parameter $ a = 0 $, as summarized in Table~\ref{tab:CCpole}.  In the hidden-charm sector, we observe that as the cutoff varies within this range, the lower poles in the $(0)0^{++}$ and $(0)1^{+-}$ systems move onto their first RSs, forming bound states. Meanwhile, the other poles tend to shift toward their respective nearby thresholds as the cutoff increases. Notably, the mass variations for the lower pole in $(0)1^{+-}$ and another pole in $(0)2^{++}$ are significant—approximately 15 MeV—due to their relatively large separation from nearby thresholds, making their dependence on the cutoff more pronounced compared to other poles.  In contrast, within the same cutoff range, all poles in the hidden-bottom sector remain on their original RSs. However, the cutoff dependence in this sector is considerably stronger than in the hidden-charm sector. In particular, the lower poles in the $(0)0^{++}$ and $(0)1^{+-}$ systems, as well as the bound-state pole in $(0)1^{++}$, exhibit mass variations ranging from 30 to 70 MeV.  In addition, the cutoff dependence of these poles on the  value of the parameter $a$, i.e. $a=0.25,0.5,0.75,1.0$, is shown in Appendix~\ref{sec:pole-pos-a}. The positions of the poles relevant to $D^*\bar D^*$ and $B^*\bar B^*$ channel thresholds are affected by the parameter $a$, but the patterns shown remain the same as the cutoff varies.     
			\begin{table*}[ht!]\centering
				\caption{Pole positions in the $D^{(*)}\bar D^{(*)} $ and $B^{(*)}\bar B^{(*)} $ systems in the isoscalar sector by varying the $\Lambda$ around $1.11$ GeV. The subscript for each entry denotes the RS where the pole is located on. The pole positions and relevant thresholds $W_i$ are given in MeV.}\label{tab:CCpole}
				\begin{ruledtabular}
				\begin{tabular}{cccccccc}
				$D^{(*)}\bar D^{(*)} $& \multicolumn{2}{c}{$(0)0^{++}$}    & $(0)1^{++}$  & \multicolumn{2}{c}{$(0)1^{+-}$}     & \multicolumn{2}{c}{$(0)2^{++}$} \\
				$W_i$ [MeV]           & $3734.5$      & $4017.1$           & $3875.8$     & $3875.8$      & $4017.1$            & \multicolumn{2}{c}{$4017.1$} \\\hline
				$\Lambda$=1.06        & $3731.2_{-+}$ & $3997.4-i0.7_{+-}$ & $3874.1_{+}$ & $3874.9_{-+}$ & $4022.5-i27.4_{+-}$ & \multicolumn{2}{c}{$4010.8-4.9_{--+}$} \\
				1.11                  & $3734.0_{-+}$ & $4000.1-i0.6_{+-}$ & $3871.7_{+}$ & $3872.8_{++}$ & $4023.3-i16.8_{+-}$ & \multicolumn{2}{c}{$4001.8-i5.3_{--+}$} \\			
				1.16                  & $3734.3_{++}$ & $4002.6-i0.5_{+-}$ & $3868.4_{+}$ & $3855.3_{++}$ & $4020.2-i10.5_{+-}$ & \multicolumn{2}{c}{$3989.2-i4.8_{--+}$} \\
				\hline\hline
				$B^{(*)}\bar B^{(*)} $& \multicolumn{2}{c}{$(0)0^{++}$}      & $(0)1^{++}$   & \multicolumn{2}{c}{$(0)1^{+-}$}      & \multicolumn{2}{c}{$(0)2^{++}$} \\
				$W_i$[MeV]            & $10559.1$      & $10649.5$           & $10604.3$     & $10604.3$       & $10649.5$          &         $10649.5$      & $10649.5$\\\hline
	            $\Lambda$=1.06 	 	  & $10512.8_{++}$ & $10648.6-i0.2_{+-}$ & $10571.3_{+}$ & $10604.2_{++}$  & $10658.5-i4.6_{+-}$& $10570.5-i0.0_{--+}$   & $10624.1-i14.4_{--+}$ \\
	            1.11 				  & $10492.5_{++}$ & $10649.2-i0.3_{+-}$ & $10561.1_{+}$ & $10599.8_{++}$  & $10657.5-i5.4_{+-}$& $10541.22 -i0.0_{--+}$ & $10622.8-i8.9_{--+}$   \\
	            1.16 				  & $10468.6_{++}$ & $10649.5-i0.1_{+-}$ & $10502.3_{+}$ & $10588.9_{++}$  & $10655.2-i4.1_{+-}$& $10533.16-i0.0_{--+}$  & $10619.6-i4.6_{--+}$  \\
					\end{tabular}
				\end{ruledtabular}
				\end{table*}

\section{Summary}\label{sec:summary}

In this work, we have systematically investigated charmonium-like and bottomonium-like hadronic molecular states near the $D^{(*)}\bar D^{(*)}$ and $B^{(*)}\bar B^{(*)}$ thresholds using the one-boson-exchange (OBE) model. Our approach respects both heavy quark spin symmetry and SU(3)-flavor symmetry. We analyzed possible near-threshold states with quantum numbers $(0)0^{++}$, $(0)1^{+\pm}$ and $(0)2^{++}$ by analytically continuing the scattering S-matrix, which was extracted from asymptotic wave functions obtained through solving the coupled-channel Schrödinger equation.

A key aspect of our analysis was investigating the role of the short-range $\delta(\bm r)$ term in the OBE model. By introducing a parameter $a$ to control its contribution, we systematically studied how this term affects the pole positions and their trajectories on different Riemann sheets. Using the well-established $X(3872)$ as a calibration point, we fixed the cutoff parameter $\Lambda$ to 1.11 GeV and explored the pole evolution as $a$ varied from 0 to 1.

For the $D^{(*)}\bar D^{(*)}$ systems in the isoscalar sector, we found: A virtual state near the $D\bar D$ threshold that could become bound under slight parameter variations; The $X(3872)$ as a $[D\bar D^*]_-$ molecular state with $(0)1^{++}$;Several near-threshold poles in the $(0)1^{+-}$ and $(0)2^{++}$ channels, with positions sensitive to the $\delta(\bm r)$ term. In the $B^{(*)}\bar B^{(*)}$ systems, we observed similar spectral patterns but with generally stronger binding, attributed to the larger reduced mass of the bottom mesons. Notable findings include: A deeply bound state about 60 MeV below the $B\bar B$ threshold; A bottomonium analogue of $X(3872)$, predicted as a bound state 30 MeV below the $B\bar B^*$ threshold; Multiple near-threshold states in various channels that warrant experimental investigation. Importantly, our analysis revealed no poles in the isovector sectors of either system within reasonable parameter ranges, suggesting that isovector molecular states are unlikely to form through OBE interactions alone.

These results provide valuable insights into the nature of exotic hadrons and offer specific predictions for future experimental searches, particularly in the bottomonium sector. The systematic treatment of the $\delta(\bm r)$ term effects also contributes to our understanding of short-range dynamics in heavy meson interactions.
    
\begin{acknowledgments}
 
 This work is supported by the Doctoral Program of Tian Chi Foundation of Xinjiang Uyghur Autonomous Region of China under grant No. 51052300506. The work of UGM was supported in part by the CAS President's International
Fellowship Initiative (PIFI) (Grant No.~2025PD0022).

\end{acknowledgments}

 \begin{appendices}
 \section{Potentials of $\pfield^{(*)} \pbfield^{(*)}$ systems}\label{sec:potential}
 We collect the potentials related to $\pfield^{(*)} \pbfield^{(*)}\to \pfield^{(*)} \pbfield^{(*)}$  scattering in the following.
 \begin{subequations}\label{eq:poten-in-p}
	\begin{itemize}
	\item $\pfield\pbfield\to \pfield \pbfield$
\begin{align}
\mathcal{V}_\sigma&=-g_s^2\frac{1}{\bm q^2+m_\sigma^2},\\
\mathcal{V}_\mathbb{V}&=-\frac{1}{2}\beta^2 g_V^2\frac{1}{\bm q^2+m_{\mathbb V}^2}.
 \end{align}
\end{itemize}
 The amplitudes for the scattering process $\pfield\pbfield\to \pfield \pbfield^*$ and $\pfield\pbfield\to \pfield^* \pbfield$ in $S$-wave are forbidden.
 \begin{itemize}
	\item $\pfield\pbfield\to \pfield^*\pbfield^*$
	\begin{align}
	   \mathcal{V}_\mathbb{V}&=-2\lambda^2 g_V^2\frac{(\bm \epsilon_3^*\times \bm q)\cdot(\bm \epsilon_4^*\times \bm q)}{\bm q^2+\mu_{\mathbb V}^2},\\
	   \mathcal{V}_{\mathbb{P}}&=\frac{g^2}{f_\pi^2}\frac{\bm\epsilon^*_3\cdot \bm q\bm\epsilon^*_4\cdot \bm q}{\bm q^2+\mu_{\mathbb P}^2}.
		\end{align}
	   \end{itemize}
 \begin{itemize}
 \item $\pfield\pbfield^*\to \pfield\pbfield^*$
 \begin{align}
	\mathcal{V}_\sigma&=-g_s^2\frac{\bm\epsilon_2\cdot \bm \epsilon_4^*}{\bm q^2+m_\sigma^2},\\
	\mathcal{V}_\mathbb{V}&=-\frac{1}{2}\beta^2g_V^2\frac{\bm\epsilon_4^*\cdot \bm\epsilon_2}{\bm q^2+m_{\mathbb V}^2}.
	 \end{align}
	\end{itemize}
 \begin{itemize}
 \item $\pfield\pbfield^*\to \pfield^*\pbfield$
 \begin{align}
	\mathcal{V}_\mathbb{V}&=-2\lambda^2 g_V^2\frac{(\bm \epsilon_3^*\times \bm q)\cdot(\bm \epsilon_2\times \bm q)}{\bm q^2+\mu_{\mathbb V}^2},\\
	\mathcal{V}_{\mathbb{P}}&=\frac{g^2}{f_\pi^2}\frac{\bm\epsilon_3^*\cdot \bm q\bm \epsilon_2\cdot \bm q}{\bm q^2+\mu_{\mathbb P}^2}.
	 \end{align}
	\end{itemize}
 \begin{itemize}
 \item $\pfield\pbfield^*\to \pfield^*\pbfield^*$
 \begin{align}
	\mathcal{V}_\mathbb{V}&=2\lambda^2 g_V^2\frac{(\bm \epsilon_3^*\times \bm q)\cdot(i\bm\epsilon_2\times \bm\epsilon_4^{*}\times \bm q) }{\bm q^2+\mu_{\mathbb V}^2},\\
	\mathcal{V}_{\mathbb{P}}&=\frac{g^2}{f_\pi^2}\frac{\bm\epsilon_3^*\cdot \bm q(i\bm\epsilon_2\times\bm\epsilon_4^*)\cdot \bm q}{\bm q^2+\mu_{\mathbb P}^2}.
	 \end{align}
	\end{itemize}
 \begin{itemize}
 \item $\pfield^*\pbfield\to \pfield^*\pbfield$
 \begin{align}
	\mathcal{V}_\sigma&=-g_s^2\frac{\bm\epsilon_1\cdot \bm \epsilon_3^*}{\bm q^2+m_\sigma^2},\\
	\mathcal{V}_\mathbb{V}&=-\frac{1}{2}\beta^2g_V^2\frac{\bm\epsilon_3^*\cdot \bm\epsilon_1}{\bm q^2+m_{\mathbb V}^2}.
	 \end{align}
	\end{itemize}
 \begin{itemize}
 \item $\pfield^*\pbfield\to \pfield^*\pbfield^*$
 \begin{align}
	\mathcal{V}_\mathbb{V}&=-2\lambda^2 g_V^2\frac{(\bm \epsilon_4^*\times \bm q)\cdot(i\bm\epsilon_1\times \bm\epsilon_3^{*}\times \bm q) }{\bm q^2+\mu_{\mathbb V}^2},\\
	\mathcal{V}_{\mathbb{P}}&=-\frac{g^2}{f_\pi^2}\frac{\bm\epsilon_4^*\cdot \bm q(i\bm\epsilon_1\times\bm\epsilon_3^*)\cdot \bm q}{\bm q^2+\mu_{\mathbb P}^2}.
	 \end{align}
	\end{itemize}
 \begin{itemize}
 \item $\pfield^*\pbfield^*\to \pfield^*\pbfield^*$
 \begin{align}
	\mathcal{V}_\sigma&=-g_s^2\frac{\bm\epsilon_1\cdot \bm \epsilon_3^* \bm\epsilon_2\cdot \bm \epsilon_4^*}{\bm q^2+m_\sigma^2},\\
	\mathcal{V}_\mathbb{V}&=-\frac{1}{2}\beta^2g_V^2\frac{\bm\epsilon_3^*\cdot \bm\epsilon_1 \bm\epsilon_4^*\cdot \bm\epsilon_2}{\bm q^2+m_{\mathbb V}^2}\notag\\
	&\quad-2\lambda^2 g_V^2\frac{(i\bm \epsilon_1\times\epsilon_3^*\times\bm q)\cdot(i\bm \epsilon_2\times\epsilon_4^*\times\bm q)}{\bm q^2+m_{\mathbb V}^2},\\
	\mathcal{V}_{\mathbb{P}}&=-\frac{g^2}{f_\pi^2}\frac{(i\bm\epsilon_1\times\bm\epsilon_3^*)\cdot \bm q (i\bm\epsilon_2\times\bm\epsilon_4^*)\cdot \bm q}{\bm q^2+m_{\mathbb P}^2}.
	 \end{align}
     where $\mu_{\rm{ex}}$ is the effective mass of the exchanged meson defined as $\mu_{\rm{ex}}^2=m_{\rm{ex}}^2-(q^0)^2$ with the energy of the exchanged meson $q^0$
\begin{align}
q^0=\frac{m_2^2-m_1^2+m_3^2-m_4^2}{2(m_3+m_4)},
\end{align}  
where $m_1(m_3)$ and $m_2(m_4)$ are the masses of the heavy and anti-heavy mesons in the initial(final) state. 
	\end{itemize}
\end{subequations}
\section{Cutoff dependence of the pole locations for $a\neq 0$}\label{sec:pole-pos-a}
In Table~\ref{tab:CCpole-other}, we present the cutoff dependence of the poles corresponding to isoscalar $J^P=0^{++},1^{+\pm},2^{++}$ states in the coupled channel analysis for $a=0.25,0.5,0.75,1.0$, comparing $D^{(*)}\bar D^{(*)}$ and $B^{(*)}\bar B^{(*)}$ systems. While both systems exhibit cutoff sensitivity, the poles in $B^{(*)}\bar B^{(*)}$ systems $-$ being farther from their respective thresholds ($W_i$) $-$ show significantly stronger dependence on $\Lambda$. Especially, for $a\geq0.75$, the higher pole in the $B^{(*)}\bar B^{(*)}$ system with $(0)2^{++}$ moves deeper into the complex energy plane (denoted by ``$\cdots$'') due to the suppression of the short-range $\delta(\bm{r})$ term contributions. The lower pole in $B^{(*)}\bar B^{(*)}$ system with $(0)1^{++}$ (binding energy $\sim$30 MeV at $\Lambda=1.11$ GeV) shows minimal $a$-dependence and represents a prime candidate for future experimental investigation.   
\begin{table*}[ht]\centering
	\caption{Pole positions in the $D^{(*)}\bar D^{(*)} $ and $B^{(*)}\bar B^{(*)} $ systems in the isoscalar sector by varying the $\Lambda$ around $1.11$ GeV when $a=0.25,0.5,0.75,1.0$. The subscript for each entry denotes the RS where the pole is located. The pole positions and relevant thresholds $W_i$  are given in MeV}\label{tab:CCpole-other}
	\begin{ruledtabular}
	\begin{tabular}{ccccccccc}
	\multirow{2}{*}{$a$}&$D^{(*)}\bar D^{(*)} $  & \multicolumn{2}{c}{$(0)0^{++}$}    & $(0)1^{++}$  & \multicolumn{2}{c}{$(0)1^{+-}$}     & \multicolumn{2}{c}{$(0)2^{++}$} \\
	&$W_i$ [MeV]                                 & $3734.5$      & $4017.1$           & $3875.8$     & $3875.8$      & $4017.1$            & \multicolumn{2}{c}{$4017.1$} \\\hline
	\multirow{3}{*}{$0.25$}&$\Lambda$=1.06       & $3730.7_{-+}$ & $4002.7-i1.0_{+-}$ & $3874.8_{+}$ & $3867.4_{-+}$ & $4020.0-i13.2_{+-}$ & \multicolumn{2}{c}{$4017.8-i3.3_{--+}$} \\
	&1.11                                        & $3733.9_{-+}$ & $4005.6-i0.9_{+-}$ & $3872.5_{+}$ & $3874.1_{-+}$ & $4023.4-i9.8_{+-}$ & \multicolumn{2}{c}{$4013.8-i5.1_{--+}$} \\			
	&1.16                                        & $3734.4_{++}$ & $4008.1-i0.8_{+-}$ & $3869.0_{+}$ & $3875.5_{++}$ & $4024.7-i7.3_{+-}$ & \multicolumn{2}{c}{$4007.6-i6.0_{--+}$} \\\hline
	\multirow{3}{*}{$0.5$}&1.06                 & $3730.3_{-+}$ & $4009.5-i1.1_{+-}$ & $3875.3_{+}$ & $3862.7_{-+}$ & $4016.8-i6.5_{+-}$ & \multicolumn{2}{c}{$4021.3-i0.8_{++-}$} \\
	&1.11                                        & $3733.8_{-+}$ & $4012.1-i1.0_{+-}$ & $3873.3_{+}$ & $3868.0_{-+}$ & $4019.0-i4.0_{+-}$ & \multicolumn{2}{c}{$4020.2-i2.0_{--+}$} \\			
	&1.16                                        & $3734.4_{++}$ & $4014.0-i0.8_{+-}$ & $3869.7_{+}$ & $3872.5_{-+}$ & $4020.2-i1.6_{+-}$ & \multicolumn{2}{c}{$4018.1-i4.2_{--+}$} \\\hline
	\multirow{3}{*}{$0.75$}&1.06                 & $3729.9_{-+}$ & $4015.7-i0.6_{+-}$ & $3875.6_{+}$ & $3862.1_{-+}$ & $4017.1-i2.0_{+-}$ & \multicolumn{2}{c}{$4022.6-i7.2_{++-}$} \\
	&1.11                                        & $3733.6_{-+}$ & $4016.9-i0.3_{+-}$ & $3873.9_{+}$ & $3866.2_{-+}$ & $4017.9-i0.4_{+-}$ & \multicolumn{2}{c}{$4023.0-i4.2_{++-}$} \\			
	&1.16                                        & $3734.4_{++}$ & $4017.3-i0.1_{-+}$ & $3870.3_{+}$ & $3870.0_{-+}$ & $4017.5-i0.9_{-+}$ & \multicolumn{2}{c}{$4023.0-i1.5_{++-}$} \\\hline
	\multirow{3}{*}{$1.0$}&1.06          		 & $3729.4_{-+}$ & $4016.2-i0.4_{-+}$ & $3875.8_{+}$ & $3865.6_{-+}$ & $4019.4-i0.6_{-+}$ & \multicolumn{2}{c}{$4023.1-i14.8_{--+}$} \\
	&1.11                 						 & $3733.5_{-+}$ & $4013.3-i1.6_{-+}$ & $3874.4_{+}$ & $3870.2_{-+}$ & $4018.9-i4.3_{-+}$ & \multicolumn{2}{c}{$4024.2-i12.2_{--+}$} \\			
	&1.16                 						 & $3734.4_{++}$ & $4008.7-i3.8_{-+}$ & $3870.8_{+}$ & $3874.0_{-+}$ & $4017.2-i9.8_{-+}$ & \multicolumn{2}{c}{$4025.0-i9.9_{--+}$} \\
	\hline\hline
	\multirow{2}{*}{$a$}&$B^{(*)}\bar B^{(*)} $  & \multicolumn{2}{c}{$(0)0^{++}$}     & $(0)1^{++}$   & \multicolumn{2}{c}{$(0)1^{+-}$}      & \multicolumn{2}{c}{$(0)2^{++}$} \\
	&$W_i$[MeV]               					 & $10559.1$     & $10649.5$           & $10604.3$     & $10604.3$       & $10649.5$          &         $10649.5$      & $10649.5$\\\hline
	\multirow{3}{*}{$0.25$}&$\Lambda$=1.06 		 & $10515.9_{++}$& $10649.4-i0.4_{+-}$ & $10576.8_{+}$ & $10600.8_{+-}$  & $10653.7-i0.8_{+-}$& $10587.2-i0.0_{--+}$   & $10623.5-i20.7_{--+}$ \\
	&1.11                						 & $10495.5_{++}$& $10649.4-i0.1_{-+}$ & $10565.0_{+}$ & $10603.8_{+-}$  & $10653.1-i3.5_{-+}$& $10561.2-i0.0_{--+}$ & $10624.0-i15.0_{--+}$   \\
	&1.16                						 & $10458.1_{++}$& $10649.1-i0.2_{-+}$ & $10551.5_{+}$ & $10601.6_{++}$  & $10652.2-i5.6_{-+}$& $10543.1-i0.0_{--+}$  & $10622.8-i9.6_{--+}$  \\\hline
	\multirow{3}{*}{$0.5$}&1.06             	 & $10518.7_{++}$& $10649.4-i0.0_{-+}$ & $10581.1_{+}$ & $10595.9_{+-}$  & $10652.6-i5.0_{-+}$& $10605.9-i0.0_{--+}$   & $10624.6-i28.0_{--+}$ \\
	&1.11                						 & $10497.7_{++}$& $10648.8-i0.0_{-+}$ & $10568.6_{+}$ & $10602.6_{+-}$  & $10649.0-i6.1_{-+}$& $10594.8-i0.0_{--+}$ & $10623.7-i23.8_{--+}$   \\
	&1.16                						 & $10462.2_{++}$& $10647.8-i0.2_{-+}$ & $10553.8_{+}$ & $10603.0_{++}$  & $10645.2-i6.3_{-+}$& $10581.0-i0.0_{--+}$  & $10624.0-i18.6_{--+}$  \\\hline
	\multirow{3}{*}{$0.75$}&1.06           		 & $10521.0_{++}$& $10647.0-i0.3_{-+}$ & $10586.0_{+}$ & $10591.1_{+-}$  & $10634.6-i5.0_{-+}$& $10615.2-i0.9_{--+}$   & $\cdots$ \\
	&1.11                 						 & $10498.5_{++}$& $10644.9-i0.3_{-+}$ & $10572.0_{+}$ & $10601.5_{+-}$  & $10628.0-i3.6_{-+}$& $10609.3-i0.1_{--+}$ & $\cdots$   \\
	&1.16                 						 & $10468.6_{++}$& $10641.9-i0.2_{-+}$ & $10556.1_{+}$ & $10603.1_{+-}$  & $10620.6-i2.6_{-+}$& $10602.4-i0.0_{--+}$  & $\cdots$  \\\hline
	\multirow{3}{*}{$1.0$}&1.06          		 & $10522.3_{++}$& $10631.7-i7.5_{-+}$ & $10589.6_{+}$ & $10586.3_{+-}$  & $10610.7-i3.1_{-+}$& $10620.1-i2.6_{--+}$   & $\cdots$ \\
	&1.11                						 & $10499.5_{++}$& $10627.2-i13.8_{-+}$& $10575.2_{+}$ & $10598.3_{+-}$  & $10606.5-i1.4_{-+}$& $10615.8-i1.3_{--+}$ & $\cdots$   \\
	&1.16          								 & $10471.6_{++}$& $10626.6-i18.6_{-+}$& $10558.4_{+}$ & $10600.1_{+-}$  & $10597.2-i2.1_{-+}$& $10611.4-i0.4_{--+}$  & $\cdots$  \\
	\end{tabular}
	\end{ruledtabular}
	\end{table*}

 \end{appendices}
 
\bibliographystyle{apsrev4-1}
\bibliography{main2.bib}

\end{document}